\documentclass[pra,floatfix,twocolumn,superscriptaddress,showpacs,preprintnumbers,longbibliography,nofootinbib]{revtex4-2}

\usepackage{soul}
\usepackage{setspace}
\usepackage{dcolumn}    
\usepackage{bm} 
\usepackage{graphicx}
\usepackage{amsmath}  
\usepackage{amsthm}   
\usepackage{enumitem}
\usepackage{latexsym}
\usepackage{amsfonts}   
\usepackage{amssymb}
\usepackage{array}      
\usepackage{epsfig}
\usepackage{braket} 
\usepackage{bbold}
\usepackage{color}
\usepackage[colorlinks=true,linkcolor=blue,urlcolor=blue,citecolor=blue,pdfusetitle]{hyperref}
\usepackage{hyperref}
\usepackage{cancel}
\usepackage{ulem}

\usepackage{tikz}

\newcommand{\eq}{Eq.~}

\DeclareRobustCommand\openzero{\leavevmode\hbox{0\kern-.55em0}}

\newcommand{\Aod}[1]{\hat{a}^\dag_{#1}}

\newcommand{\gs}{\tikz[baseline , yshift=2pt]{\node[draw=black, fill=green, minimum size=8pt, anchor=base] {}; }}
\newcommand{\ws}{\tikz[baseline, yshift=2pt]{\node[draw=black, fill=white, minimum size=8pt, anchor=base] {}; }}

\date{\today}

\begin{document}

\author{Dario Cilluffo}
\email{dario.cillufo@uni-ulm.de}
\author{Matthias Kost} \author{Martin B. Plenio}

\affiliation{Institute of Theoretical Physics, Ulm University, Albert-Einstein-Allee 11, 89081 Ulm, Germany}
\affiliation{Center for Integrated Quantum Science and Technology (IQST), 89081 Ulm, Germany}

\begin{abstract}
We develop a Heisenberg-picture tensor-network formulation of collision-free Gaussian Boson Sampling, providing a direct Fock-space expression for output probabilities in terms of experimentally accessible quantities. The resulting representation naturally recovers the Hafnian structure while revealing the decomposition of GBS probability into a phase-insensitive contribution and a hierarchy of interference sectors associated with pairs of perfect matchings. 
As an application, we investigate phase diffusion and show how it progressively suppresses many-body interference, driving the output statistics toward a classical dimer-model regime. Our results establish a transparent framework for connecting experimentally characterized phase fluctuations with the loss of quantum interference in photonic quantum sampling experiments.\end{abstract}

\title{Phase-noise induced many-body interference suppression in Gaussian Boson Sampling}
\maketitle

Gaussian states play a central role across quantum physics from continuous-variable quantum optics to condensed matter, and are central to bosonic quantum computing platforms such as Scattershot (SBS) \cite{doi:10.1126/sciadv.1400255} and Gaussian Boson Sampling (GBS) \cite{PhysRevLett.119.170501,PhysRevA.100.032326,PhysRevLett.113.100502}. These platforms, together with standard Boson Sampling (AABS) \cite{AA}, provide near-term experimental evidence of quantum advantage.
Although Boson Sampling currently provides the most solid evidence of quantum advantage, all existing experiments are affected by experimental imperfections, with photon loss being the most significant. Estimating the impact of these imperfections is important not only from a theoretical perspective—since proofs of quantum advantage are strictly valid only in the idealized limit—but also from an experimental standpoint, as it provides a way to assess the level of control achievable in passive photonic platforms \cite{BulmerSAA}. This will become increasingly crucial in the future, particularly in the context of progressing toward universal optical quantum computing.
Tensor networks \cite{Orus_2019,montangero2018introduction,Schollwoeck_2011,lacroix2026} have gained increasing visibility in this field, with the main aim of clarifying the boundaries of quantum advantage in bosonic sampling platforms \cite{cilluffo2024,PhysRevA.108.052604}, thereby refining and clarifying the range of physically relevant applications of these systems \cite{Oh_2021,oh2024classical,PhysRevA.108.052604}, including recent studies of phase-noisy GBS based on matrix-product-operator representations \cite{Paryzkova2025}.
Based on the Heisenberg-picture tensor-network formalism for bosonic dynamics introduced in~\cite{CilluffoOBMPS}, in this Letter we provide an alternative derivation of GBS probabilities as Hafnians to that of Ref.~\cite{PhysRevA.100.032326}, which naturally resolves the output probability into distinct interference sectors and provides a direct framework to investigate their sensitivity to experimental imperfections. In particular, we study phase instability in time-integrated measurements, where averaging over fluctuating Gaussian configurations generally produces a non-Gaussian mixture whose photon-number statistics cannot be reconstructed from a single covariance matrix.
Motivated by recent experiments showing different responses of nonclassicality witnesses to phase noise in GBS and SBS \cite{stefszky2025}, we show how phase diffusion affects the Hafnian probability by progressively suppressing interference between its perfect-matching contributions, ultimately reducing it to a phase-insensitive Hafnian of non-negative matrix elements. Importantly, increasing the photon number does not counteract this effect, revealing a trade-off between system size and the phase stability required to preserve many-body interference in GBS.
\\
\textit{Heisenberg picture tensor network representation.}
A pure 
$M$-mode bosonic Gaussian state $|\mathcal{G}\rangle$ is the ground state of a quadratic Hamiltonian in the canonical variables
$(x,p)$ \cite{Serafini2023,RevModPhys.84.621}.
Equivalently, up to a phase-space displacement, 
$|\mathcal{G}\rangle$ can be written as a squeezed-vacuum state, i.e. as the action of an exponential of a quadratic form in the creation operators on the multimode vacuum \cite{liu2025}:
\begin{align}
|\mathcal{G} \rangle = \frac{1}{\mathcal{N}}\exp\left\{ \frac{1}{2}\sum_{i,j=1}^M T_{ij} \Aod{i} \Aod{j} \right\} |\mathbb{0}\rangle\,,
\label{eq:state}
\end{align}
where $\mathcal{N}=\det(\mathbb{1}- T^\dag T)^{-1/4}$\cite{balian1969nonunitary} and $T=(U^\dag)^{-1}V^\dag$ is the symmetric pairing-amplitude matrix, with the $M\times M$ complex matrices $U$ and $V$ fulfilling the bosonic symplectic conditions ($U^\dag U - V^\dag V = \mathbb{1}$ , $V^\dag U^* - U^\dag V^* =0$). For the specific case relevant to Gaussian Boson Sampling, the input state is a product of two–mode squeezed vacuum states. In this case the pairing matrix 
$T$ takes the block-diagonal form
 \begin{align}
     T= \bigoplus_{i=0}^{M/2-1} \theta_i  ~\sigma_x
 \end{align}
 where $\theta_i=-e^{i \phi_{2i,2i+1}} \tanh{(\xi_{2i,2i+1})}$, $\phi_{2i,2i+1}$ and $\xi_{2i,2i+1}$ are the squeezing phase and amplitude of the $i$th pair of input modes respectively, and $\sigma_x$ is the Pauli matrix.
 After evolution through a unitary transformation $\mathcal{U}$, we get, for the quadratic form
\begin{align}
\mathcal{U} (\sum_{ij} T_{ij} \Aod{i} \Aod{j})  & 
= \sum_{i j} \tau_{ij} \Aod{i} \Aod{j} \,,
\end{align}
where
\begin{align}
    {\tau}_{ij}&=(\mathcal{U}^T T \mathcal{U})_{ij}=
\sum_{m=0}^{M/2-1} \theta_m \langle 1_i 1_j |\mathcal{U}|1_{2m} 1_{2m+1}\rangle  
\,.
\label{eq:tau}
\end{align}
The symmetric matrix $\tau$ generally contains nonzero diagonal entries, corresponding to two photons occupying the same output mode. Since we restrict to collision-free detection events, these terms do not contribute. We therefore retain only the off-diagonal terms and restrict the sum to $i>j$ to avoid double counting.
We can now represent this sum as a product of augmented vectors like
\begin{align}
\mathcal{U} (\sum_{i>j} T_{ij} \Aod{i} \Aod{j}) & = \prod_{i>j} 
\left( \begin{matrix} \mathbb{1} & \mathbb{0} \\ \tau_{ij} \Aod{i}\Aod{j} & \mathbb{1} \end{matrix}\right) 
\,,
\label{eq:matrix_notation}
\end{align}
where, with a slight abuse of notation, throughout the following we identify such auxiliary-space products with their lower-left matrix element, which contains the physical operator of interest.
The operator can be expressed in OBMPS form as \cite{CilluffoOBMPS}
\begin{align}
\mathcal{U} (\sum_{i>j} T_{ij} \Aod{i} \Aod{j})  &= \sum_{\sigma_{ij}\in\{0,1\}} \prod_{i>j} A^{[ij]}_{\sigma_{ij}}  \bigotimes_{kl}(\Aod{k}\Aod{l})^{\sigma_{kl}} 
\,,
\label{eq:single}
\end{align}
where
$    A^{[ij]}_0= \mathbb{1}_2, ~~~ A^{[ij]}_1= \tau_{ij}\sigma_-\,$.
Thus each term of the expansion \eq\eqref{eq:state} is a product of $m$ sums \eqref{eq:single}.
The quantity of interest in gaussian boson sampling is  $ \langle \mathbf{n}| \mathcal {U} |\mathcal{G}\rangle$, where $|\mathbf{n}\rangle$ is a $n$-photon Fock state over $M$ modes, which reduces to the expression
\begin{align}
\langle \mathbf{n}|\mathcal{U} |\mathcal{G} \rangle = \frac{1}{ \mathcal{N} (n/2)!} \langle \mathbf{n} |  \mathcal{U} (\sum_{i>j} T_{ij} \Aod{i} \Aod{j} )^{n/2} |\mathbb{0}\rangle \,,
\end{align}
which can be rewritten in OBMPS as
\begin{align}
\langle \mathbf{n} | \mathcal{U} |\mathcal{G} \rangle & = \frac{1}{\mathcal{N}(n/2)!} \sum_{  \rho\in {\rm PM}({\bar{i}}) } \prod_{(ij)\in \rho} \mathcal{A}^{[ij]\, n_{ij}}_{}\,. 
\label{eq:prob1}
\end{align}
where $\bar{i}(\mathbf{n}):= \bar{i}$ denotes the set of mode indices corresponding to the occupied modes in the Fock state $|\mathbf{n}\rangle$, ${\rm PM}(\bar{i})$ are the perfect matchings among the elements in $\bar{i}$ and
\\
\begin{align}
\mathcal{A}^{[ij]\, n_{ij}}  = \!\!\!  \sum_{\sum_{k=1}^{n/2} \sigma^{}_k = n_{ij}} \!\!
A^{[ij]}_{\,\sigma^{}_1} \otimes \ldots \otimes A^{[ij]}_{ \,\sigma^{}_{n/2}}
\,.
\label{eq:expansion}
\end{align}
Note that, so far, we have allowed the possibility of repeated pairings; in this case, the vector $\bar{i}$ may contain multiple copies of the same index. We now restrict to the collision-free regime, where each index appears at most once $(n_{ij}=0,1)$, and in this setting, the above term can be represented as the graph
\begin{align}
\mathcal{A}_{}^{[ij]\, 1} = [ \, \gs ~ \underbrace{\ws \ldots \ws}_{n/2-1}\, ]\,,
\end{align}
where 
$
    \ws\ \: :=A^{[ij]}_{0} = \mathbb{1}_2, ~ \gs:=A^{[ij]}_{1}\,,
$
the horizontal alignment denotes a tensor product
and the square brackets indicate the sum over all possible placements of the green boxes \cite{CilluffoOBMPS}.
Accordingly, each product of $\mathcal{A}$-matrices in \eq
\eqref{eq:prob1} reads
\begin{align}
~ \left.
\begin{matrix}
[\, \gs ~ \ws ~ \ws \ldots \ws \,]\\
[\, \ws ~ \gs ~ \ws \ldots \ws \,]\\
\ldots \\
[\, \ws ~ \ws ~ \ws  \ldots \gs \,]\\
\end{matrix} ~~\right\} n/2\, ,
\label{eq:fake_comb}
\end{align}
where vertical alignments denotes ordinary matrix products. Note that, due to the nilpotency of $A^{[ij]}_{1}$, vertical alignments of green boxes yield zero.
Unlike in Fock Boson Sampling case described in \cite{CilluffoOBMPS}, each $A$-matrix does not carry any information about the specific input port of the photons: as a consequence, all $(n/2)!$ valid placements of the $n/2$ green boxes contribute identically. Their multiplicity exactly cancels the prefactor $1/(n/2)!$ arising from the exponential expansion in Eq.~(9), so that it is sufficient to retain a single representative placement. As a result, we get
\begin{align}
\langle \mathbf{n} | \mathcal{U} | \mathcal{G} \rangle & = \frac{1}{\mathcal{N}} \sum_{  \rho\in {\rm PM}({\bar{i}}) } \prod_{(ij)\in \rho} \tau_{ij}\,, 
\label{eq:prob2}
\end{align}
which matches exactly the definition of the Hafnian: the sum ranges over the set of perfect matchings within the set $\bar{i}$.
This provides a compact derivation of the collision-free GBS Hafnian amplitude directly in Fock space from the OBMPS representation.
\\
\textit{Decomposition of probability.} Eq.~\eqref{eq:prob2} provides a direct connection between the probability and the experimentally relevant parameters (squeezing and transmittivities), without resorting to the covariance-matrix formalism. This is particularly convenient when analyzing the effect of defects on the complexity of computing probabilities, which underlies the computational hardness of GBS.
Furthermore, this framework allows one to straightforwardly extend the analysis beyond Gaussian defects.
In this section we will focus on the effects of phase noise and losses.
Phase noise specifically perturbs the relative phases of the contributing terms in the products in \eq\eqref{eq:prob2}. In terms of $\tilde{\tau}=\tau/\mathcal{N}^{2/n}$, the probability of observing a specific Fock state $|\mathbf{n}\rangle$ reads
\begin{align}
|\langle \mathbf{n} |\mathcal{U}|\mathcal{G} \rangle|^2 & = \sum_{\chi,\rho\in {\rm PM}({\bar{i}})}\;
\prod_{\substack{(i,j)\in \chi\\(k,l)\in \rho}}  \tilde{\tau}_{ij}
\tilde{\tau}_{kl}^{*}
~=\sum_{k=0}^{n/2} I_k(\bar{i})
\label{eq:prob3}
\end{align}
where we defined the \textit{interference sector} $I_k(\bar{i})$ as
\begin{align}
 I_k (\bar{i})= \sum_{\substack{\chi,\rho\in {\rm PM}({\bar{i}})\\ |\chi \Delta \rho|/2=k}}\;
\prod_{\substack{(i,j)\in \chi\\(l,m)\in \rho}}  \tilde{\tau}_{ij}
\tilde{\tau}_{lm}^{*}
\end{align}
For $k=0$ we get
\begin{align}
I_0 (\bar{i})=\sum_{ \chi \in {\rm PM}({\bar{i}}) }\;
\prod_{(i,j)\in \chi} \tilde{\tau}_{ij}
\prod_{(k,l)\in \chi}
\tilde{\tau}_{kl}^{*}
= {\rm Haf}({|\tilde{\tau}(\bar{i})|^2})\,,
\end{align}
where the notation $\tilde{\tau}(\bar{i})$ indicates the submatrix of $\tilde{\tau}$ obtained by restricting its rows and columns to the indices contained in $\bar{i}$.
The next contribution is identically zero. Indeed, any two perfect matchings on the same vertex set $\bar{i}$ differ in at least two edges, so no pair of matchings can have symmetric difference of size one. Therefore, $I_1 (\bar{i}) = 0$.
The higher order contributions are grouped in the \textit{interference term} $\mathfrak{I}$, which is defined as
\begin{align}
    \mathfrak{I}(\tilde{\tau}(\bar{i})) &: 
= \sum_{k\geq 2} I_k (\bar{i})
 \label{eq:ulmian}
\end{align}

\textit{Phase-noise induced many-body interference suppression.}
Phase noise can occur due to source instability (fluctuations of
$\arg(\theta_m)$ in Eq.~\eqref{eq:tau}), as well as due to fluctuations of
the couplings and interferometer phases (fluctuations of the phase of the
two-photon amplitude terms in Eq.~\eqref{eq:tau}). While interferometric
instabilities can in principle be reduced by stabilization techniques,
source fluctuations are generally more difficult to eliminate. In general, fluctuations of the microscopic parameters entering
Eq.~\eqref{eq:tau} induce both amplitude and phase fluctuations of the
resulting matrix elements $\tau_{ij}$.
In the following, we focus on phase
fluctuations and model the matrix elements as
\begin{equation}
    \tau_{ij}(t)=|\tau_{ij}|e^{-i \phi_{ij}(t)}
    \,
    \label{eq:deftauph}
\end{equation}
with $\phi_{ij}(t)$ undergoing some unspecified stochastic process, neglecting the fluctuations of the modulus.
This approximation is motivated by the fact that amplitude fluctuations
modify the instantaneous values of the Hafnian matrix elements, but do not
directly generate a systematic loss of coherence between the different
components contributing to the interference sector. In contrast,
phase fluctuations accumulate between different terms of the squared Hafnian
expansion and lead, under time averaging, to a suppression of the coherent
interference contributions. Therefore, the effective phase-noise model
captures the component of the fluctuations responsible for the dephasing-induced suppression effect we investigate here.
It is useful to consider separately the different contributions arising from the decomposition of the output probability. The terms belonging to the interference sector $\mathfrak{I}$ exhibit an explicit dependence on the phase factors, and in particular on the phases of the squeezed sources. As we will show, time-dependent phase fluctuations can modulate these contributions and progressively suppress them through dephasing. For this reason, the interference sector will be the main focus of this section.
The contribution $I_0$ contains no interference between distinct perfect-matching amplitudes and is therefore insensitive to the relative phases responsible for cancellations in the interference contribution. It can be interpreted as the partition function $Z_D(\bar{i})$ of a classical weighted dimer model with non-negative edge weights \cite{Fisher1966Dimer,10.1007/978-981-15-0294-1_2}. Consequently, the resulting ensemble is not covered by the complex-Gaussian random-matrix assumptions used in standard GBS hardness arguments, although exact hafnian evaluation remains generally difficult.
For strictly positive matrices satisfying $\delta\leq \tau_{ij} \leq1$, with fixed $\delta>0$, Barvinok gives a deterministic quasipolynomial-time relative approximation for the hafnian \cite{barvinok2017approximatingpermanentshafnians}. 
More generally, for finite-precision non-negative inputs, Stockmeyer-type
approximate counting places the problem of multiplicatively approximating
${\rm Haf}(A)$ in $\mathrm{FBPP}^{\mathrm{NP}}$ \cite{Stockmeyer1983}. Remarkably, the same linear non-negative Hafnian weighting that characterizes the fully dephased contribution, $I_0(\bar{i})$, appears in the quantum-inspired construction of Oh \textit{et al.} \cite{PRXQuantum.5.020341}. Their construction combines independent two-photon sampling processes to generate a distribution proportional to ${\rm Haf}(A(\bar{i}))$, with $A$ real, symmetric, and entrywise non-negative, without requiring the many-body interference underlying GBS. Thus, complete suppression of $\mathfrak{I}$ leaves the same Hafnian-weighted probability structure as an efficiently classically samplable construction. This provides a direct computational interpretation of the dephasing limit and shows that the standard GBS quantum-advantage arguments based on coherent many-body interference no longer apply.
We now move on to the terms in $\mathfrak{I}$. In photonic experiments, the output statistics are accumulated over a finite interval $T$, which may
correspond either to the physical detector response window or to the acquisition time over which photon-counting statistics are collected.
Consequently, in the presence of time-dependent phase fluctuations, the observed probability corresponds to a time- and ensemble-averaged quantity rather than to the instantaneous probability in Eq.~\eqref{eq:prob2}.
From the definition of $\tau_{ij}$ in \eqref{eq:deftauph}, the product of the two perfect-matching amplitudes entering the interference contribution between $\chi$ and $\rho$ acquires a relative phase. Since phases associated with common edges cancel, this is given by
\begin{align}
\Delta\Phi_{\chi\rho}(t)
=
\sum_{(ij)\in\chi\setminus\rho}\delta\phi_{ij}(t)
-
\sum_{(ij)\in\rho\setminus\chi}\delta\phi_{ij}(t)\,,
\end{align}
where $\delta\phi_{ij}(t)$ denotes the stochastic fluctuation around the static
phase $\phi_{ij}^{(0)}$.
To quantify the suppression of the interference between the two perfect-matching contributions, we define the time-averaged coherence factor
\begin{align}
w_{\chi\rho}(T)
=
\frac{1}{T}\int_0^T \! dt\;
e^{-i\Delta\Phi_{\chi\rho}(t)}
.
\label{eq:Gamma}
\end{align}
In general, $w_{\chi\rho}$ depends on the specific pair of perfect
matchings and on the correlations between the fluctuating phases. For
zero-mean jointly Gaussian fluctuations, it is completely determined by
their covariance matrix $C(t)$ through
\begin{align}
\mathbb{E}\!\left[
e^{-i\Delta\Phi_{\chi\rho}(t)}
\right]
=
\exp\!\left[
-\frac{1}{2}
\mathbf{c}_{\chi\rho}^{T}
C(t)
\mathbf{c}_{\chi\rho}
\right],
\end{align}
where $\mathbf{c}_{\chi\rho}$ has entries $+1$, $-1$, or $0$ depending on
whether an edge belongs only to $\chi$, only to $\rho$, or to both.

We now specialize to a homogeneous effective dephasing model in which the
relative phase associated with each of the $k$ mismatches is represented by
independent Wiener processes $\phi_a(t)$, with
$\mathrm{Cov}[\phi_a(t),\phi_b(t')]
=2\sigma^2\delta_{ab}\min(t,t')$ and
$\mathrm{Var}[\phi_a(t)]=2\sigma^2t$, such that
\begin{align}
\Delta\Phi_{\chi\rho}(t)=\sum_{a=1}^{k}\phi_a(t).
\end{align}
In this case the coherence factor depends only on the mismatch index $k$,
and we denote it by $w_k$. Since
$S_k(t)=\sum_{a=1}^{k}\phi_a(t)$ is Gaussian with
$\mathrm{Var}[S_k(t)]=2k\sigma^2t$, we obtain
\begin{align}
\mathbb{E}\!\left[
\cos S_k(t)
\right]
=
e^{-k\sigma^2t}.
\end{align}
Hence
\begin{align}
w_k
=
\frac{1}{T}\int_0^T dt\,e^{-k\sigma^2t}
=
\frac{1-e^{-k\sigma^2T}}{k\sigma^2T}.
\label{eq:w1}
\end{align}
Thus, within this homogeneous model, interference sectors with larger
mismatch are progressively more strongly suppressed.

Under the same assumptions, its variance is given by (see the Supplementary Material for details): 
\begin{align}
var(w_k)
=&
\frac{1}{T^2}
\left[
\frac{k\sigma^2 T - 1 + e^{-k\sigma^2 T}}{(k\sigma^2)^2} \notag
\right.\\&\left.+
\frac{3 + e^{- 4 k\sigma^2 T} - 4 e^{- k\sigma^2 T}}{12 k^2\sigma^4}
\right]
- w_k^2 \,.
\end{align}
In light of the discussion above we can recast \eq\eqref{eq:prob3} as
\begin{align}
    |\langle \mathbf{n} |\mathcal{U}|\mathcal{G} \rangle|^2 & 
~={Z}_D(\bar{i}) + \mathfrak{I}_w\,,
\label{eq:decomp}
\end{align}
with
\begin{align}
    \mathfrak{I}_w = \sum_{k=2}^{n/2} w_k I_k(\bar{i})\,.
\end{align}
\begin{figure*}
\centering
\includegraphics[scale=0.7,angle=0]{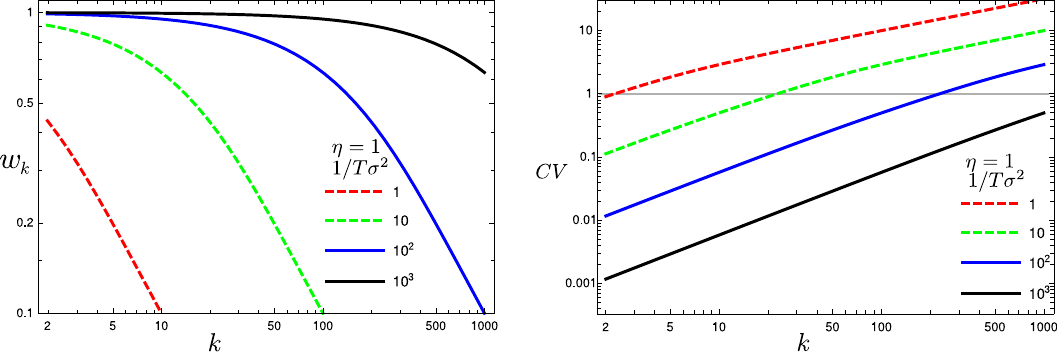}	
\caption{Weighting factor $w_k$ (left) for lossless setup and coefficient of variation (CV) as functions of the number $k$ for different values of the ratio between the detector integration rate and the phase diffusion rate.}
\label{fig:fig1}
\end{figure*}

\textit{Uniform losses.}
In addition, our framework allows us to estimate the effect of losses on the interference contribution $\mathfrak{I}$. We first consider losses at the level of a fixed photon-number sector. Uniform losses can be described by assigning to each mode an amplitude transmission factor $\sqrt{\eta}$, where $\eta=1-\epsilon$ is the single-photon transmission probability. Consequently, the pair-correlation matrix transforms as
$\tau \rightarrow \tau_{\eta}=\eta\tau$,
so that an $n$-photon contribution acquires the expected survival factor
$w_k \rightarrow \eta^n w_k$.
Since photon loss acts independently of the phase diffusion process, the attenuation associated with losses and the phase-averaging factors factorize within each fixed photon-number sector. Therefore, the loss-induced modification of a given interference contribution preserves the same phase-noise dependence discussed above, with an additional multiplicative factor determined by the probability of the corresponding loss event.
A complete description of an unconditional lossy Gaussian Boson Sampling experiment requires accounting for the redistribution of photon-number sectors induced by losses: an observed $n$-photon event may originate from higher-order input sectors in which photons are lost. Applying the loss channel to the density matrix of the evolved state, $\mathcal{U}|\mathcal{G}\rangle\langle\mathcal{G}|\mathcal{U}^\dagger$, and projecting onto the detected state $|\mathbf n\rangle$, the lossy probability can be written as
\begin{align}
|\langle \mathbf{n}|\mathcal{U}|\mathcal{G}\rangle|^2_{\mathcal L}=
\sum_{\mathbf p,\mathbf q}
{\rm Haf}(\tilde{\tau}(\bar p))
{\rm Haf}(\tilde{\tau}(\bar q))^*
\langle\mathbf n|
\mathcal L
\left(
|\mathbf p\rangle\langle\mathbf q|
\right)
|\mathbf n\rangle ,
\end{align}
where $\bar p(\mathbf p)$ and $\bar q(\mathbf q)$ denote the repeated-index lists associated with the corresponding Fock states, and
$
\mathcal L(\rho)=\sum_{\mathbf{m}} K_\mathbf{m}\rho K_\mathbf{m}^\dagger
$
is the loss channel, with the vector $\mathbf{m}$  specifying the number of photons lost in each mode. 
For uniform losses, the same Kraus operator acts on the ket and bra components of the density matrix.
For a given loss pattern $m$, the Kraus operator maps a Fock state according to
$
K_{\mathbf m}|\mathbf p\rangle\propto |\mathbf p-\mathbf m\rangle .
$
Therefore,
$\langle\mathbf n|K_{\mathbf m}|\mathbf p\rangle\neq0
$
only if
$n_i=p_i-m_i$ for every mode $i$. The corresponding contribution from the bra component requires simultaneously
$n_i=q_i-m_i$.
Since the same loss pattern acts on both sides of the density matrix, non-vanishing terms necessarily satisfy
$p_i=q_i~\forall i$,
and hence $\mathbf p=\mathbf q$. Uniform losses, therefore, produce an incoherent weighted sum over sectors with different photon numbers.
Each squared Hafnian contribution can thus be decomposed into a phase-insensitive part and an interference part, with the latter suppressed by phase diffusion as discussed above. Consequently, the effect of losses enters only through the sector weights, and the characterization of phase-noise-induced interference suppression can be performed sector by sector. 
Since each fixed photon-number sector admits the same decomposition into classical and interference contributions, characterizing the suppression of the interference weight within a sector is sufficient to understand the effect of phase noise throughout the lossy mixture.

\textit{Discussion.}
We note that phase noise and losses affect the two sectors of
Eq.~\eqref{eq:decomp} in qualitatively different ways. Phase diffusion
selectively suppresses the interference contribution $\mathfrak{I}$, while
leaving the incoherent contribution $I_0$ unchanged within the effective
phase-noise model considered here. Although contributions to $\mathfrak{I}$
with large mismatch number $k$ are typically small in absolute magnitude,
they correspond to higher-order interference terms in the Hafnian expansion,
whose number grows combinatorially with photon number. Since the suppression
increases with $k$, phase noise progressively attenuates these contributions,
effectively reducing the interference structure until, in the complete
dephasing limit, only the incoherent contribution remains. An experiment with
$n/2$ photon pairs contains interference sectors with mismatch up to $k=n/2$;
increasing the photon number therefore opens higher-order sectors that are progressively more susceptible to phase diffusion. Uniform losses, on the other hand, rescale the contributions within a fixed photon-number sector by the same overall factor. Within our model they therefore attenuate their absolute magnitude without affecting the relative phase-noise dependence or the coefficient of variation of $w_k$.
In Fig.~\ref{fig:fig1}, we show $w_k$ and the coefficient of variation $CV=\sqrt{\operatorname{var}(w_k)}/w_k$ for different values of
$1/(\sigma^2T)$ in the lossless regime ($\eta=1$). We observe a smooth crossover from near-full coherence to a regime in which the interference weights decay with increasing $k$. At the same time, the CV increases and
eventually crosses $CV=1$, marking a fluctuation-dominated regime in which the standard deviation becomes comparable to the mean. Within the noise model
considered here, this occurs only at sufficiently large $k$ in the parameter regimes of interest. The combined effect of phase diffusion and uniform
losses, including representative high-loss regimes, is discussed in the Supplemental Material.

\textit{Conclusion.}
Our results have two main consequences for Gaussian Boson Sampling experiments.
First, preserving many-body interference requires the detection window to
remain sufficiently short compared with the characteristic phase-fluctuation
timescale. As phase fluctuations accumulate, many-body interference is progressively suppressed, driving the output statistics away from the regime underlying standard GBS quantum-advantage arguments.
Second, our analysis reveals a trade-off between photon number and phase
stability. Increasing the number of detected photons enlarges the many-body
interference structure of the Hafnian, but simultaneously opens interference
sectors with larger mismatch index $k$, which are more strongly affected by
phase diffusion. This behavior is qualitatively reminiscent of recent experiments \cite{stefszky2025}, where degradation of nonclassicality witnesses based on accumulated photon-counting statistics was observed in the presence of phase noise and losses in GBS, while no analogous effect was found in SBS. The latter behavior is naturally reproduced by our model, since in SBS the phase factors enter the relevant amplitudes only as global phases and therefore cancel from the output probabilities. While the microscopic origin of the experimental fluctuations differs from the effective dephasing model considered here, both results highlight the sensitivity of many-photon interference to phase instability.
The Wiener model adopted here provides an analytically tractable reference
for this dephasing mechanism. In realistic implementations, fluctuations of the squeezed-light sources and optical paths may induce correlated,
non-Wiener phase dynamics as well as amplitude fluctuations. Once these noise
processes are experimentally characterized, for example through measured phase time series or correlation functions, they can be included through
the corresponding stochastic interference weights, providing a route toward experiment-specific characterizations of interference degradation.
\\
\\
\textit{Acknowledgements.} We acknowledge T. Haas, M. Stefszky and G. Lo Monaco for fruitful discussions.
This research was supported by the BMBF project PhoQuant (Grant No. 13N16110).

\bibliography{biblio}

\clearpage

\clearpage
\onecolumngrid
\setcounter{equation}{0}
\setcounter{figure}{0}

\begin{center}
    \textbf{\large Supplemental Material for}\\[0.5em]
    \textbf{\large ``Phase-noise induced many-body interference suppression in Gaussian Boson Sampling''}\\[1em]
    Dario Cilluffo,\textsuperscript{1,2}
Matthias Kost,\textsuperscript{1,2}
and Martin B. Plenio\textsuperscript{1,2}\\[0.5em]

\textsuperscript{1}\textit{Institute of Theoretical Physics, Ulm University,
Albert-Einstein-Allee 11, 89081 Ulm, Germany}\\
\textsuperscript{2}\textit{Center for Integrated Quantum Science and Technology (IQST),
89081 Ulm, Germany}\\[0.5em]
\end{center}

\section{ Variance of the time-averaged phase weight}
\label{app:var}

We compute the variance of the time-averaged quantity
\begin{align}
w_k = \frac{1}{T}\int_0^T dt \; \cos\big(S(t)\big), \qquad 
S(t)=\sum_{a=1}^k \phi_a(t) \,,
\end{align}
where $\phi_a(t)$ are independent Wiener processes with diffusion rate $\sigma^2$, such that $\mathrm{Var}[\phi_a(t)] = 2\sigma^2 t$. It follows that $S(t)$ is a Gaussian process with
\begin{align}
\mathrm{Var}[S(t)] = 2k\sigma^2 t \,.
\end{align}

The variance of $w$ is given by
\begin{align}
\mathrm{Var}(w_k) = \mathbb{E}[w_k^2] - \mathbb{E}[w_k]^2 \,,
\end{align}
with
\begin{align}
\mathbb{E}[w_k^2] =
\frac{1}{T^2} \int_0^T dt \int_0^T dt'\;
\mathbb{E}\big[\cos(S(t))\cos(S(t'))\big] \,.
\end{align}

Using $\cos A \cos B = \tfrac{1}{2}[\cos(A-B)+\cos(A+B)]$, we obtain
\begin{align}
\mathbb{E}[\cos(S(t))\cos(S(t'))]
&=
\frac{1}{2}\Big(
\mathbb{E}[\cos(S(t)-S(t'))]
+
\mathbb{E}[\cos(S(t)+S(t'))]
\Big) \,.
\end{align}

Since $S(t)$ is Gaussian, we use $\mathbb{E}[\cos X] = \exp(-\tfrac{1}{2}\mathrm{Var}[X])$. The covariance of Wiener processes is
\begin{align}
\mathrm{Cov}(\phi(t),\phi(t')) = 2\sigma^2 \min(t,t') \,,
\end{align}
which implies
\begin{align}
\mathrm{Cov}[S(t),S(t')] = 2k\sigma^2 \min(t,t') \,.
\end{align}

\paragraph{Difference term.}
\begin{align}
\mathrm{Var}[S(t)-S(t')] &= \mathrm{Var}[S(t)] + \mathrm{Var}[S(t')] - 2\,\mathrm{Cov}[S(t),S(t')] \notag\\
&= 2k\sigma^2 (t+t' - 2\min(t,t')) \notag\\
&= 2k\sigma^2 |t-t'| \,,
\end{align}
so that
\begin{align}
\mathbb{E}[\cos(S(t)-S(t'))] = e^{-k\sigma^2 |t-t'|} \,.
\end{align}

\paragraph{Sum term.}
\begin{align}
\mathrm{Var}[S(t)+S(t')] &= \mathrm{Var}[S(t)] + \mathrm{Var}[S(t')] + 2\,\mathrm{Cov}[S(t),S(t')] \notag\\
&= 2k\sigma^2 (t+t' + 2\min(t,t')) \,,
\end{align}
leading to
\begin{align}
\mathbb{E}[\cos(S(t)+S(t'))] = e^{-k\sigma^2 (t+t' + 2\min(t,t'))} \,.
\end{align}

Combining both contributions,
\begin{align}
\mathbb{E}[\cos(S(t))\cos(S(t'))]
=\!\!
\frac{1}{2}\! \left(
e^{-k\sigma^2 |t-t'|}
\!+\!
e^{-k\sigma^2 (t+t' + 2\min(t,t'))}\!
\right).
\end{align}

Evaluating the double integral, one obtains
\begin{align}
\mathbb{E}[w_k^2]
&=
\frac{1}{T^2}
\left[
\frac{1}{(k\sigma^2)^2}\Big(k\sigma^2 T - 1 + e^{-k\sigma^2 T}\Big)
\right.\\&\left.+
\frac{1}{12 k^2\sigma^4}
\Big(3 + e^{- 4 k\sigma^2 T} - 4 e^{- k\sigma^2 T})\Big)
\right] \,.
\end{align}

Finally, subtracting $\mathbb{E}[w_k]^2$, we obtain 
\begin{align}
\mathrm{Var}(w_k)
&=
\frac{1}{T^2}\!\!
\left[
\frac{(k\sigma^2 T - 1 + e^{-k\sigma^2 T}) }{(k\sigma^2)^2}
+
\frac{3 + e^{- 4 k\sigma^2 T} - 4 e^{- k\sigma^2 T}}{12 k^2\sigma^4}
\right]
\!\!- w_k^2 \,,
\end{align}
where
\begin{align}
w_k = \frac{1- e^{- k \sigma^2 T}}{k \sigma^2 T} \,.
\end{align}

\section{Effect of uniform losses on mean and variance}
\label{app:loss}
We model uniform losses by assuming that each photon is transmitted
independently with probability $\eta$. For a sector containing $m$ pairs,
the survival of all $2m$ photons introduces an overall transmission factor
$\eta^{2m}$. Assuming that loss is independent of the phase-noise-induced
fluctuations, the loss-affected weight can therefore be written as
\begin{align}
w_{\mathrm{loss}}=\eta^{2m}w.
\end{align}
Its mean and second moment scale as
\begin{align}
\mathbb{E}[w_{\mathrm{loss}}]
&=\eta^{2m}\mathbb{E}[w],\\
\mathbb{E}[w_{\mathrm{loss}}^2]
&=\eta^{4m}\mathbb{E}[w^2],
\end{align}
and consequently
\begin{align}
\operatorname{Var}(w_{\mathrm{loss}})
=\eta^{4m}\operatorname{Var}(w).
\end{align}
The standard deviation therefore scales with the same factor $\eta^{2m}$
as the mean, so that the coefficient of variation is invariant under
uniform losses,
\begin{align}
CV_{\mathrm{loss}}
=
\frac{\sqrt{\operatorname{Var}(w_{\mathrm{loss}})}}
{\mathbb{E}[w_{\mathrm{loss}}]}
=
CV.
\end{align}

Uniform losses attenuate the absolute magnitude of the interference
contributions while preserving their relative fluctuations, as quantified by
the loss-independent coefficient of variation derived above. For a sector
containing $m$ photon pairs, uniform transmission $\eta$ rescales all
interference contributions by the same factor $\eta^{2m}$, corresponding to
a relative suppression of $1-\eta^{2m}$, independently of the mismatch
index $k$. For instance, at $50\%$ transmission this corresponds to a
suppression of $75\%$, $93.75\%$, and $99.90\%$ for $m=1$, $2$, and $5$,
respectively. Thus, uniform loss does not modify the relative hierarchy of
the interference sectors, but rapidly reduces their absolute magnitude as
the photon number increases.

\end{document}